\pdfoutput=1

\documentclass[11pt]{article}
\usepackage[final]{acl}

\usepackage{listings}
\usepackage{times}
\usepackage{latexsym}
\usepackage{float}
\usepackage{graphicx}   
\usepackage{subcaption}  
\usepackage[T1]{fontenc}
\usepackage{graphicx}
\usepackage{color}
\usepackage{lineno}
\usepackage{booktabs}
\usepackage{makecell}
\usepackage{amsfonts} 
\usepackage{multirow}
\usepackage[utf8]{inputenc}

\usepackage{microtype}

\usepackage{inconsolata}
\usepackage[utf8]{inputenc}
\usepackage{graphicx}

\title{CLaMP 2: Multimodal Music Information Retrieval \\ Across 101 Languages Using Large Language Models}

\author{
    {\makebox[0.18\textwidth]{\textbf{Shangda Wu}} \hfill
    \makebox[0.18\textwidth]{\textbf{Yashan Wang}} \hfill
    \makebox[0.18\textwidth]{\textbf{Ruibin Yuan}} \hfill
    \makebox[0.18\textwidth]{\textbf{Zhancheng Guo}} \hfill
    \makebox[0.18\textwidth]{\textbf{Xu Tan}}}
    \\[0.5ex]
    \makebox[0.18\textwidth]{\textbf{Ge Zhang}} \hfill
    \makebox[0.18\textwidth]{\textbf{Monan Zhou}} \hfill
    \makebox[0.18\textwidth]{\textbf{Jing Chen}} \hfill
    \makebox[0.18\textwidth]{\textbf{Xuefeng Mu}} \hfill
    \makebox[0.18\textwidth]{\textbf{Yuejie Gao}}
    \\[0.5ex]
    \makebox[0.18\textwidth]{\textbf{Yuanliang Dong}} \hfill
    \makebox[0.18\textwidth]{\textbf{Jiafeng Liu}} \hfill
    \makebox[0.18\textwidth]{\textbf{Xiaobing Li}} \hfill
    \makebox[0.18\textwidth]{\textbf{Feng Yu}} \hfill
    \makebox[0.18\textwidth]{\textbf{Maosong Sun}}
    \\[1ex]
    {\small Details of contributors, correspondence, and affiliations are on Page 9}
    \\[1.5ex]
    \url{https://github.com/sanderwood/clamp2}}

\begin{document}
\maketitle
\begin{abstract}
Challenges in managing linguistic diversity and integrating various musical modalities are faced by current music information retrieval systems. These limitations reduce their effectiveness in a global, multimodal music environment. To address these issues, we introduce CLaMP 2, a system compatible with 101 languages that supports both ABC notation (a text-based musical notation format) and MIDI (Musical Instrument Digital Interface) for music information retrieval. CLaMP 2, pre-trained on 1.5 million ABC-MIDI-text triplets, includes a multilingual text encoder and a multimodal music encoder aligned via contrastive learning. By leveraging large language models, we obtain refined and consistent multilingual descriptions at scale, significantly reducing textual noise and balancing language distribution. Our experiments show that CLaMP 2 achieves state-of-the-art results in both multilingual semantic search and music classification across modalities, thus establishing a new standard for inclusive and global music information retrieval.
\end{abstract}

\section{Introduction}
As a cross-cultural art form that transcends geographical boundaries, music is being accessed globally more than ever, as people seek diverse content to enhance their aesthetic experience. However, current Music Information Retrieval (MIR) systems struggle to meet this demand, particularly in the area of multilingual retrieval. For example, a Japanese user searching for \textit{"Brazilian choro music with themes of celebration and carefreeness"} in their native language may face significant challenges. Keyword-based retrieval methods might return choro music, but they often fail to capture the specific themes the user is searching for. Meanwhile, existing cross-modal MIR models remain heavily focused on English \cite{DBLP:conf/ismir/HuangJLGLE22,DBLP:conf/icassp/ElizaldeDIW23,DBLP:conf/icassp/DohWCN23a}, making effective multilingual semantic search challenging.

A key limitation in the development of multilingual MIR systems is that most music-text datasets are predominantly in English \cite{DBLP:journals/corr/abs-2301-11325,DBLP:conf/nips/LanzendorferGFW23,manco2023thesong}. As a result, MIR models struggle to process text queries in non-English languages. Additionally, textual noise—such as inconsistent metadata and variations in terminology—complicates the task of matching descriptions to the appropriate music. Addressing these challenges requires advanced techniques to manage multilingual data more effectively and reduce noise, allowing MIR systems to bridge linguistic and aesthetic gaps.

Recent Large Language Models (LLMs) \cite{DBLP:journals/corr/abs-2303-08774,dubey2024llama3herdmodels,DBLP:journals/corr/abs-2403-05530} have demonstrated robust performance in language-related tasks. LLMs have been used in previous cross-modal MIR models and music-text dataset curation to generate coherent descriptions and annotations \cite{DBLP:conf/ismir/DohCLN23,DBLP:conf/icassp/WuCZHBD23,DBLP:journals/corr/abs-2306-00110,DBLP:journals/corr/abs-2406-02255}. This has proven effective in improving text quality and enhancing model performance. Since LLMs are typically multilingual, they hold significant potential for generating high-quality music descriptions in multiple languages. This could overcome the limitations of current MIR systems and significantly enhance global music accessibility.

To leverage these advancements, we introduce CLaMP 2, a cross-modal MIR model designed to effectively link multilingual text with diverse music data. The model includes a text encoder \cite{DBLP:conf/acl/ConneauKGCWGGOZ20} and a music encoder \cite{DBLP:conf/ismir/WuY0S23}, which are aligned by contrastive learning \cite{DBLP:conf/nips/Sohn16,DBLP:journals/corr/abs-1807-03748}. Pre-trained on a substantial dataset of 1.5 million ABC-MIDI-text triplets, CLaMP 2 incorporates LLM-generated text to boost its multilingual processing capabilities. This enables the model to gain a deep understanding of musical concepts and their subtleties across various languages. Notably, CLaMP 2 supports 101 languages and unifies two symbolic music formats—ABC notation and MIDI—with new encoding methods into one framework. By enhancing multilingual semantic search and integrating diverse music data, CLaMP 2 sets a new standard for global MIR, enabling users to access music from a wide range of linguistic and cultural contexts.

The contributions of this paper are as follows:

\begin{itemize}
    \item We utilized GPT-4 \cite{DBLP:journals/corr/abs-2303-08774} to refine the multilingual corpus used for contrastive learning. This reduced noise, balanced language distribution, and improved the overall quality of the pre-training dataset.
    \item We enhanced an existing music encoder \cite{DBLP:conf/ismir/WuY0S23} to support both ABC notation and MIDI data using novel encoding techniques for better musical representation. Empirical results prove that joint training on both modalities enhances extracted feature quality.
    \item CLaMP 2 achieves state-of-the-art results in multiple MIR tasks, showing that LLM-generated data significantly boosts multilingual retrieval performance.
\end{itemize}

\section{Related Work}
\subsection{Multilingual Language Models}
Multilingual Language Models (MLMs), trained on text from various languages, play a crucial role in Natural Language Processing (NLP) and related fields. Early MLM research used word embeddings to represent words of different languages in a shared representation space. For instance, fastText \cite{DBLP:conf/eacl/GraveMJB17} provided pre-trained word embeddings for multilingual NLP tasks, enabling the calculation of cross-language similarities.

In recent years, more advanced MLMs based on complex neural network architectures \cite{DBLP:conf/nips/VaswaniSPUJGKP17} have been introduced. Examples include mBERT\footnote{\url{https://github.com/google-research/bert/blob/master/multilingual.md}}, mBART \cite{DBLP:journals/tacl/LiuGGLEGLZ20}, and mT5 \cite{DBLP:conf/naacl/XueCRKASBR21}, all of which evolved from their monolingual counterparts \cite{DBLP:conf/naacl/DevlinCLT19,DBLP:conf/acl/LewisLGGMLSZ20,DBLP:journals/jmlr/RaffelSRLNMZLL20} and are well-suited to multilingual environments. XLM-R \cite{DBLP:conf/acl/ConneauKGCWGGOZ20} has shown strong performance in low-resource languages, demonstrating the efficacy of large-scale multilingual modeling. In contrast to English-centric models, M2M-100 \cite{DBLP:journals/jmlr/FanBSMEGBCWCGBL21} allows direct translation between 100 languages, marking a major step forward in multilingual translation. Additionally, SeamlessM4T \cite{DBLP:journals/corr/abs-2308-11596} overcomes the limitations of traditional translation models by supporting up to 100 languages and enabling translation between speech and text, as well as within the same modality, all in a unified framework.

Lately, LLMs \cite{DBLP:journals/corr/abs-2406-12793,DBLP:journals/corr/abs-2401-04088,DBLP:journals/corr/abs-2407-10671} have become increasingly multilingual to better serve a global audience. By utilizing diverse linguistic data from large training corpora, LLMs have improved both their accessibility and usefulness for users around the world. Similarly, cross-modal MIR systems must evolve to support multilingual queries, enabling more inclusive retrieval and interaction across languages.

\subsection{Applications of LLMs in Music}
Recent advancements in LLMs have greatly influenced the music field. Specifically, many models and datasets now leverage LLM-generated text to improve both music understanding and generation.

MuseCoco \cite{DBLP:journals/corr/abs-2306-00110} uses LLMs to translate musical attributes into coherent, detailed descriptions, enabling more precise control over music generation. Similarly, Noise2Music \cite{DBLP:journals/corr/abs-2302-03917} leverages pre-trained LLMs to generate musical descriptions paired with audio data, enriching the dataset with semantically rich captions. Beyond generative models, TTMR++ \cite{DBLP:conf/icassp/DohLJN24} enhances text-to-music retrieval by incorporating detailed descriptions from a fine-tuned LLaMA 2 \cite{DBLP:journals/corr/abs-2307-09288} model alongside metadata, leading to more relevant and accurate search results. For dataset curation, MidiCaps \cite{DBLP:journals/corr/abs-2406-02255} provides over 168 thousand MIDI files, each paired with detailed musical attributes like tempo, key, and instrumentation. These attributes are then utilized by Claude 3 Opus \cite{anthropic2024claude3} to generate fluent captions for the MIDI files. LP-MusicCaps \cite{DBLP:conf/ismir/DohCLN23} employs GPT-3.5 Turbo \cite{DBLP:conf/nips/Ouyang0JAWMZASR22} to generate music descriptions and explores different instructions to create diverse captions, resulting in 2.2 million captions and 0.5 million audio clips.

Nevertheless, the aforementioned efforts mainly focus on improving text coherence and fluency and are English-exclusive. To the best of our knowledge, CLaMP 2 is the first to leverage the multilingual capabilities of LLMs to improve multilingual performance in the music field.

\begin{figure*}[t]
    \centering
    \includegraphics[width=\textwidth]{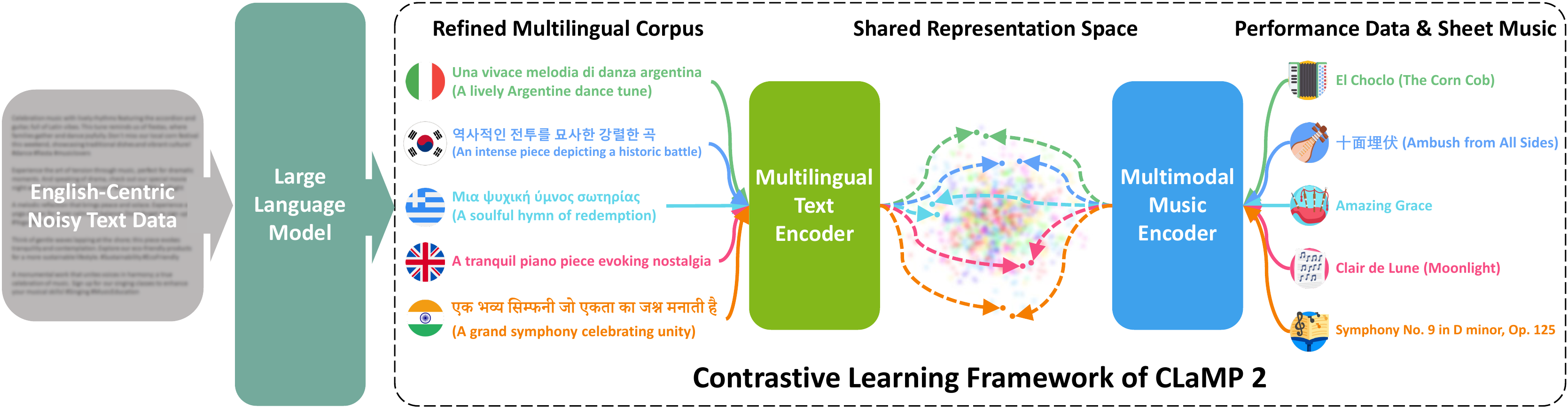}
    \caption{CLaMP 2 is a cross-modal MIR model that uses contrastive learning to link multilingual text and multimodal music data. It employs GPT-4 to refine the multilingual corpus, reducing noise and achieving a more balanced language distribution. The refined text data is then encoded by a multilingual text encoder. Meanwhile, music data in both ABC notation (sheet music) and MIDI (performance data) formats is processed by a multimodal music encoder. Both encoders project data into a shared representation space to connect text and music.}
    \vspace{-1em}
\end{figure*}

\section{CLaMP 2}
In this section, we present the CLaMP 2 framework. We begin with an overview of contrastive learning for modality alignment, followed by discussions of the multilingual text and multimodal music encoders. Finally, we introduce the data sources used for pre-training and elaborate on how we leverage GPT-4 to enhance data quality.

\subsection{Contrastive Learning}
Contrastive learning \cite{DBLP:conf/nips/Sohn16, DBLP:journals/corr/abs-1807-03748} is a powerful technique in various applications for aligning different modalities \cite{DBLP:conf/icml/RadfordKHRGASAM21,DBLP:conf/cvpr/GirdharELSAJM23}. It minimizes the distance between paired representations and maximizes that for unpaired ones. This effectively maps semantically related features (e.g., an image and its caption) close together in a shared representation space while separating unrelated ones.

As shown in Fig. 1, CLaMP 2 applies contrastive learning to ABC-MIDI-text triplets. The music encoder processes both ABC notation and MIDI data, while the text encoder handles the corresponding text inputs. During each training epoch, either ABC or MIDI data from each triplet is randomly selected for the music encoder, while the text encoder processes either the original metadata or the refined multilingual descriptions generated by GPT-4. Additionally, instrument information is removed from the music data 90\% of the time, encouraging the model to focus on broader musical concepts rather than specific instrumentations. Both encoders project data into a shared representation space to learn the underlying connections between music and text. In this space, similar musical and textual concepts are clustered together, while dissimilar ones are kept apart.

\subsection{Multilingual Text Encoder}
CLaMP 2 uses XLM-R-base \cite{DBLP:conf/acl/ConneauKGCWGGOZ20}, a multilingual text encoder based on RoBERTa \cite{DBLP:journals/corr/abs-1907-11692}. With 270 million parameters, it is pre-trained on a 2.5TB cleaned CommonCrawl corpus that spans a wide range of languages, enabling it to capture diverse linguistic nuances.

During each training epoch, the input text for each triplet is randomly selected with the following probabilities: 50\% for the raw text data, 25\% for LLM-generated English descriptions, and 25\% for LLM-generated non-English descriptions. This selection ensures a balanced exposure to both real-world and LLM-generated multilingual data. Additionally, we apply text dropout from the original CLaMP framework \cite{DBLP:conf/ismir/WuY0S23} to the raw text data. It helps the model generalize better by reducing overfitting to specific input patterns.

For computational efficiency, we set the maximum text length to 128. Longer texts are truncated in one of three ways with equal probability: using the first, the last, or randomly selecting a segment of 128 tokens. This minimizes bias that could arise from relying on a single truncation method.

\subsection{Multimodal Music Encoder}
CLaMP 2's multimodal music encoder supports multi-track music encoding in both ABC notation and MIDI. Although they can be mutually converted, they are different in nature. ABC notation (sheet music), a text-based sheet music representation like stave notation, is theory-oriented and ideal for presenting complex musical concepts to musicians for study and analysis. In contrast, MIDI (performance data) precisely encodes performance information related to timing and dynamics, thus suitable for music production and live performance.

The music encoder of CLaMP 2 is built on M3 \cite{DBLP:conf/ismir/WuY0S23}, a self-supervised model designed for feature extraction from sheet music based on bar patching. This method divides sheet music into bar-like segments, maintaining musical coherence while improving efficiency. M3 has an asymmetric encoder-decoder framework: the patch-level encoder extracts contextualized features from patches, while the char-level decoder then uses these features to autoregressively reconstruct each corresponding bar. During pre-training, 45\% of patches are randomly selected and uniformly processed with corresponding probabilities: 80\% masked, 10\% shuffled, and 10\% unchanged. M3 is optimized via cross-entropy loss to predict original patches from noisy input.

Compared to the previous M3 model, we made several important improvements to CLaMP 2's multimodal music encoder. Notably, it now supports MIDI data. MIDI messages are first read losslessly from the original file using the mido library\footnote{\url{https://github.com/mido/mido}} and then converted to the MIDI Text Format (MTF) proposed in this paper. As MTF is a text-based format, each message read from it can be treated as a patch for M3. It offers two main advantages: 1) seamless integration with the M3 framework, enabling the same training methods, and 2) lossless MIDI-to-MTF conversion, which preserves all information and avoids common quantization errors found in existing MIDI representations \cite{DBLP:journals/nca/OoreSDES20,DBLP:conf/mm/HuangY20,DBLP:conf/aaai/HsiaoLYY21}.

Another improvement is restructuring ABC notation into a voice-interleaved form. As previous research has verified \cite{DBLP:journals/corr/abs-2404-06393}, this can significantly reduce the difficulty of modeling multi-track ABC notation and is conducive to training. Importantly, our implementation of interleaved ABC notation adheres to syntax rules, ensuring compatibility with existing ABC notation tools.

The patch-level encoder is expanded to 12 layers to better capture complex musical features, while the char-level decoder remains at 3 layers, both with a hidden size of 768. Each patch can hold up to 64 characters, and with a maximum of 512 patches per input sequence, M3 can support a total input of 32,768 characters. Longer sequences are truncated by randomly selecting 512 patches from the start, middle, or end, with equal probability.

For details on interleaved ABC notation and MTF, please see Appendix A and B, respectively.

\begin{figure}[t]
    \centering
    \includegraphics[width=0.475\textwidth]{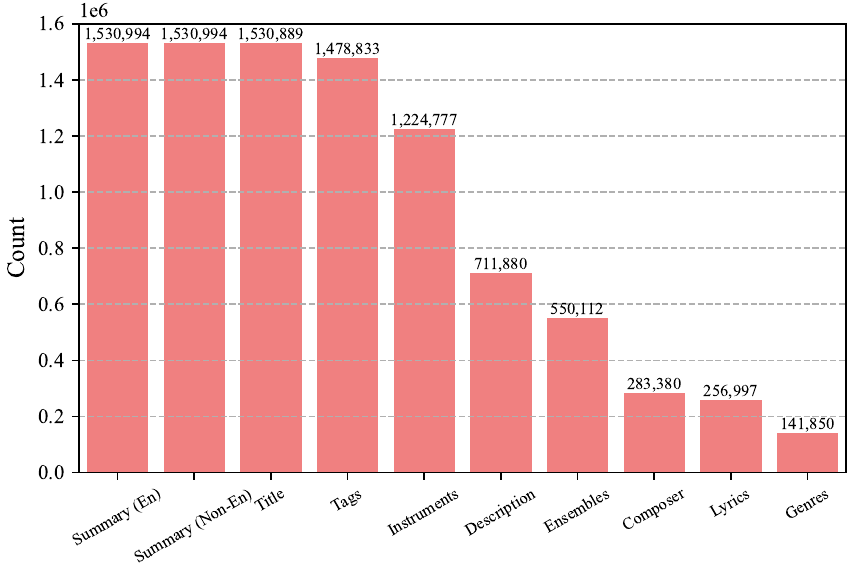}
    \vspace{-0.5em}
    \caption{The distribution of counts for different text types within the LLM-processed pre-training dataset.}
    \vspace{-1em}
\end{figure}

\subsection{Data Sources}
The pre-training dataset for both M3 and CLaMP 2 comes from two sources: the Million MIDI Dataset (MMD) \cite{DBLP:conf/acl/ZengTWJQL21} and the WebMusicText (WebMT) dataset \cite{DBLP:conf/ismir/WuY0S23}. They cover various music genres, such as popular and classical, from single- to multi-track compositions.

The MMD consists of over 1.5 million MIDI files, compiled by crawling a vast collection of music files and filtering out any malformed or blank entries. On the other hand, the WebMT dataset, comprising 1.4 million music-text pairs, includes formats like MusicXML, LilyPond, and ABC notation. These were standardized into ABC notation following an initial conversion to MusicXML. To prevent information leakage, natural language elements were removed from the ABC files.

To unify the datasets, we convert MMD to ABC, WebMT to MIDI, and merge them to get 3 million ABC-MIDI-text triplets. Admittedly, converting MMD to ABC may lead to the loss of performance details, and converting WebMT to MIDI may result in the loss of certain score-related information. Nevertheless, the key benefit is that it enriches data diversity, thus enhancing the model's ability to generalize across different musical modalities.

However, variations in text quality pose significant challenges. A substantial amount of non-musical content in the text data diminishes the effectiveness of pre-training by introducing noise that detracts from relevant musical information. Furthermore, as Fig. 3 shows, the dataset has an imbalanced language distribution (detected by the langid library\footnote{\url{https://github.com/saffsd/langid.py}}): English accounts for two-thirds of the data, while most languages contribute less than 1MB. This imbalance restricts the model's ability to effectively link music with various languages.

\begin{figure*}[t]
    \centering
    \includegraphics[width=\textwidth]{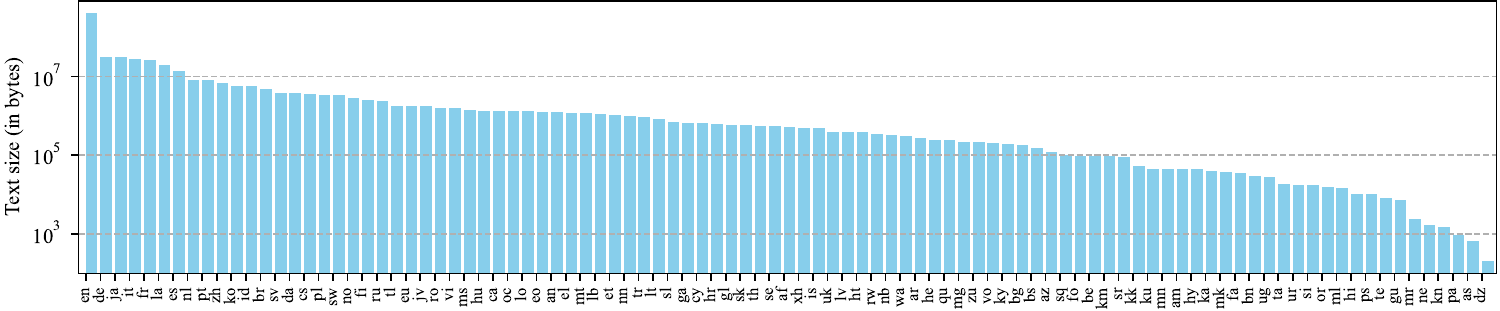}
    \caption{The amount of data for 97 languages found in the original metadata, displayed in order of magnitude.}
\end{figure*}

\begin{figure*}[t]
    \centering
    \includegraphics[width=\textwidth]{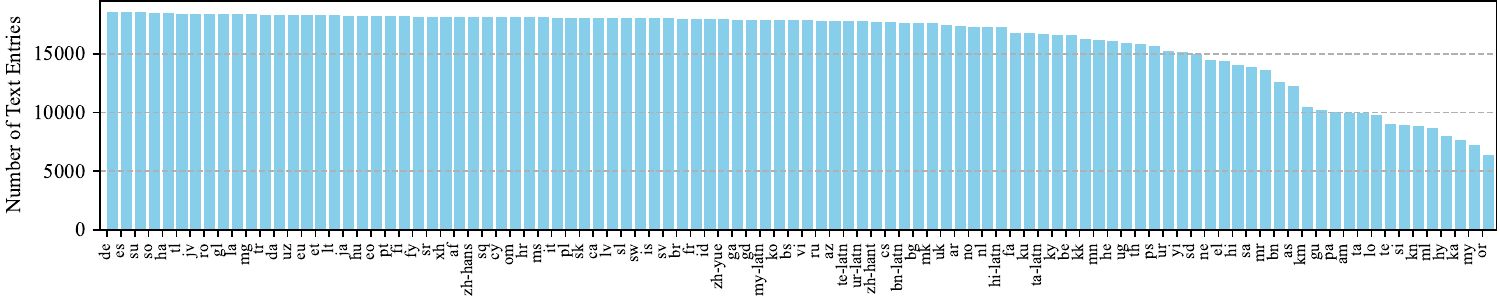}
    \caption{Count of text entries for 100 non-English languages generated by GPT-4.}
    \vspace{-1em}
\end{figure*}

\subsection{LLM-Based Metadata Processing}
To improve text quality and mitigate imbalanced language distribution, we employed GPT-4 \cite{DBLP:journals/corr/abs-2303-08774} to filter and enrich the text data. The prompt given to GPT-4 consisted of a system instruction along with two examples illustrating the desired outputs, which enhanced its understanding of our requirements. GPT-4 was tasked with identifying relevant music-related elements in each entry, and subsequently generating concise summaries in multiple languages based on these elements.

Entries were excluded for lacking specific musical details, containing vague comments like \textit{"this is a good song,"} or having no significant relation to the music. For valid entries, GPT-4 generated concise summaries in English and a randomly selected non-English language from the 100 languages tokenizable by XLM-R. However, some responses in low-resource languages did not conform to the expected format, resulting in fewer entries for these languages. Nevertheless, as shown in Fig. 4, GPT-4 significantly enhanced language balance, resulting in a total of 1.6 million ABC-MIDI-text triplets.

As our dataset is derived from two sources, duplicate entries may occur. To resolve this, we merged triplets with identical components, resulting in 1.5 million unique triplets—approximately 1.3 million from WebMT and 0.2 million from MMD.

GPT-4 cleaned the pre-training dataset and enriched it with multilingual descriptions in 101 languages. This significantly enhanced CLaMP 2’s multilingual MIR capabilities. Details on the prompt and text examples are in Appendix C.

\section{Experiments}
\subsection{Settings}
We evaluated the proposed models, M3 and CLaMP 2, on music classification and semantic search tasks. Training both models together on 8 NVIDIA H800 GPUs took approximately 800 hours. We split the data, allocating 99\% for training and 1\% for validation. The models were trained for up to 100 epochs. We adopted mixed-precision acceleration \cite{DBLP:conf/iclr/MicikeviciusNAD18} to enhance training efficiency. The AdamW optimizer \cite{DBLP:conf/iclr/LoshchilovH19} was utilized, along with a 1,000-step warm-up \cite{DBLP:journals/corr/GoyalDGNWKTJH17}.

For M3, the batch size was set to 128, and the learning rate was 1e-4. For CLaMP 2, initialized from M3's patch-level encoder and XLM-R, the batch size was set to 1024, the learning rate was 5e-5, and the logit scale was set to 1.

The ablation study for M3 and CLaMP 2 included several variants. For M3, three configurations were examined to assess the impact of mixing musical modalities on performance: M3-ABC, trained only on ABC data; M3-MIDI, trained only on MIDI data; and full M3, trained on both. For CLaMP 2, five ablations were carried out to understand the contribution of different text data sources: CLaMP 2 (w/o en), excluding LLM-generated English data; CLaMP 2 (w/o nen), excluding LLM-generated non-English data; CLaMP 2 (w/o meta), excluding the original raw text data; CLaMP 2 (w/o LLM), excluding all LLM-generated data; and the full CLaMP 2 setup, using all available text data.

\begin{table*}[t!]
  \begin{center}
    \caption{Classification performance for ABC notation and MIDI was assessed across three datasets: WikiMT (1,010 pieces, 8 genres), VGMIDI (204 pieces, 4 emotions), and Pianist8 (411 pieces, 8 composers). Underlined values indicate the top M3 model, while bold values denote the overall best performance among all models.}
    \resizebox{0.95\linewidth}{!}{
      \begin{tabular}{p{1.7cm} p{1.5cm}<{\centering} p{1.7cm}<{\centering} p{1.7cm}<{\centering} p{1.7cm}<{\centering} p{1.7cm}<{\centering} p{1.7cm}<{\centering} p{1.7cm}<{\centering} p{1.7cm}<{\centering}}
        \toprule 
        \multirow{2}{*}{\centering\textit{Model}} & \multirow{2}{*}{\centering\textit{Modality}} & \multicolumn{2}{c}{\centering\textit{WikiMT}} & \multicolumn{2}{c}{\centering\textit{VGMIDI}} & \multicolumn{2}{c}{\centering\textit{Pianist8}} \\ 
        \cline{3-8} 
         &  & F1-macro & Accuracy & F1-macro & Accuracy & F1-macro & Accuracy \\
        \midrule 
        \textit{M3-MIDI} & MIDI & 0.2586 & 0.4158 & 0.4700 & 0.5854 & 0.8683 & 0.8674 \\
        \textit{M3-ABC} & ABC & 0.2416 & 0.4010 & 0.4955 & 0.6098 & 0.7339 & 0.7470 \\
        \textit{M3} & MIDI & \underline{0.2621} & \underline{0.4257} & 0.5399 & 0.6098 & \underline{\textbf{0.9199}} & \underline{\textbf{0.9157}} \\
        \textit{M3} & ABC & 0.2349 & 0.4010 & \underline{0.6016} & \underline{0.6341} & 0.7395 & 0.7590 \\
        \midrule 
        \midrule 
        \textit{MusicBERT} & MIDI & 0.1746 & 0.3219 & 0.5127 & 0.5850 & 0.8379 & 0.8413 \\
        \textit{CLaMP} & ABC & 0.3452 & 0.4267 & 0.6453 & 0.6866 & 0.7067 & 0.7152 \\
        \textit{CLaMP 2} & MIDI & 0.2898 & 0.4455 & 0.5246 & 0.6585 & 0.8927 & 0.8916 \\
        \textit{CLaMP 2} & ABC & \textbf{0.3990} & \textbf{0.4653} & \textbf{0.7449} & \textbf{0.8049} & 0.8025 & 0.8072 \\
        \bottomrule 
      \end{tabular}%
    }
  \end{center}
  \vspace{-1em}
\end{table*}

\subsection{Music Classification Across Modalities}
This evaluation assesses the classification capabilities of various models across three datasets, each highlighting a specific aspect of music.

\begin{itemize}
    \item \textbf{WikiMT} \cite{DBLP:conf/ismir/WuY0S23}: It contains 1,010 lead sheets in ABC notation from Wikifonia\footnote{\url{http://www.synthzone.com/files/Wikifonia/Wikifonia.zip}}, labeled with 8 genre classes according to the relevant Wikipedia entries.

    \item \textbf{VGMIDI} \cite{DBLP:conf/ismir/FerreiraW19}: It contains 204 MIDI scores from video game soundtracks, annotated with 4 emotion classes based on valence and arousal levels.

    \item \textbf{Pianist8} \cite{DBLP:journals/corr/abs-2107-05223}: It includes 411 piano performances automatically transcribed from audio to performance MIDI \cite{DBLP:journals/taslp/KongLSWW21} and labeled with 8 composer styles.
\end{itemize}

We evaluated our model against state-of-the-art baselines in symbolic music understanding.

\begin{itemize}
\item \textbf{CLaMP} \cite{DBLP:conf/ismir/WuY0S23}: A cross-modal MIR model designed to connect text and sheet music. It is pre-trained on WebMT using bar patching and masked music modeling.
\item \textbf{MusicBERT} \cite{DBLP:conf/acl/ZengTWJQL21}: A self-supervised MIR model for representation learning, pre-trained on MMD through OctupleMIDI encoding and bar-level masking.
\end{itemize}

In this evaluation, we utilized only the representations from the music encoder. Given that the text encoder was not involved, the multilingual capabilities were not investigated. As a result, CLaMP 2 under evaluation included all available text data.

Notably, we employed a linear classifier on the top layer of each model to assess the quality of musical representations. We evaluated each benchmark in both MIDI and ABC formats to analyze how the models utilize information from different musical modalities.

The results in Table 1 indicate that mixing musical modalities significantly benefits M3. When trained with both ABC and MIDI, M3 outperformed its single-modality counterparts on all benchmarks. This implies that training with ABC and MIDI together improves its feature extraction capability for both modalities.

Despite being pre-trained on only 0.2 million native MIDI pieces, M3 consistently outperformed MusicBERT in MIDI classification tasks. This performance advantage is attributed to our proposed MTF, which preserves all MIDI information during text conversion. In contrast, MusicBERT's OctupleMIDI encoding suffers from information loss, which weakens its performance.

Once aligned with text data, CLaMP 2 generally outperforms M3 across benchmarks, though performance varies by modalities. In ABC notation, CLaMP 2 achieves top accuracies of 0.4653 and 0.8049 in WikiMT and VGMIDI, respectively, both of which emphasize score information. However, in Pianist8, which focuses on performance details, CLaMP 2 excels in MIDI with an accuracy of 0.8916, a significant improvement over the original CLaMP. Still, this falls slightly below M3’s 0.9157, likely due to limited performance MIDI data in the pre-training dataset. This shortage may have caused a slight decline after contrastive learning. Despite this, CLaMP 2 remains highly effective across musical modalities, showing its strong potential for music classification.

\begin{table*}[t!]
  \begin{center}
    \caption{The semantic search performance of CLaMP 2 across the WikiMT and MidiCaps benchmarks under diverse experimental settings. Both datasets contain texts exclusively in English.}
    \resizebox{\linewidth}{!}{
      \begin{tabular}{p{3.2cm} p{1.4cm}<{\centering} p{1.4cm}<{\centering} p{1.4cm}<{\centering} p{1.4cm}<{\centering} p{1.4cm}<{\centering} p{1.4cm}<{\centering} p{1.4cm}<{\centering} p{1.4cm}<{\centering}}
        \toprule 
        \multirow{2}{*}{\textit{Setting}} & \multicolumn{4}{c}{\textit{WikiMT (1,010 ABC-text pairs)}} & \multicolumn{4}{c}{\textit{MidiCaps (1,010 MIDI-text pairs)}} \\
        \cline{2-9} 
         & MRR & HR@1 & HR@10 & HR@100 & MRR & HR@1 & HR@10 & HR@100 \\
        \midrule 
        \textit{CLaMP 2} & \textbf{0.3438} & \textbf{0.2705} & 0.4870 & \textbf{0.7956} & 0.2695 & 0.1653 & 0.4782 & \textbf{0.8634} \\
        \textit{CLaMP 2 (w/o en)} & 0.3234 & 0.2455 & 0.4800 & 0.7846 & 0.2708 & 0.1723 & 0.4752 & 0.8436 \\
        \textit{CLaMP 2 (w/o nen)} & 0.3359 & 0.2615 & \textbf{0.4880} & 0.7735 & 0.2490 & 0.1574 & 0.4158 & 0.8297 \\
        \textit{CLaMP 2 (w/o meta)} & 0.2856 & 0.2104 & 0.4218 & 0.7585 & 0.1940 & 0.1050 & 0.3713 & 0.7901 \\
        \textit{CLaMP 2 (w/o LLM)} & 0.2797 & 0.2094 & 0.4068 & 0.7375 & \textbf{0.2772} & \textbf{0.1762} & \textbf{0.4822} & 0.8614 \\
        \textit{CLaMP} & 0.2561 & 0.1931 & 0.3693 & 0.7020 & 0.1236 & 0.0666 & 0.2416 & 0.6412 \\
        \bottomrule 
      \end{tabular}%
    }
  \end{center}
\end{table*}

\subsection{Semantic Search on Native English Data}
Benchmarks in symbolic MIR are relatively scarce. To the best of our knowledge, WikiMT \cite{DBLP:conf/ismir/WuY0S23} and MidiCaps \cite{DBLP:journals/corr/abs-2406-02255} are the only two publicly available music-text datasets for symbolic music. WikiMT pairs 1,010 ABC notation pieces with Wikipedia text, focusing on cultural and historical context. MidiCaps, built on the Lakh MIDI dataset \cite{DBLP:phd/us/Raffel16}, includes 168,407 pairs with descriptions of musical features like tempo and chord progression. These datasets have different focuses: WikiMT emphasizes cultural-context understanding, while MidiCaps targets musical feature analysis.

As the pre-training data includes the Lakh MIDI dataset (a subset of MMD), we took precautions to prevent data leakage. To this end, we randomly selected 1,010 pieces from the MidiCaps validation set to match the size of WikiMT, which contains only non-training data. CLaMP 2 uses the original formats for testing on these benchmarks. Because the original CLaMP does not support MIDI, we converted the MidiCaps data into ABC notation for its evaluation.

\begin{figure}[t]
    \centering
    \begin{subfigure}{0.455\textwidth}
        \centering
        \includegraphics[width=\textwidth]{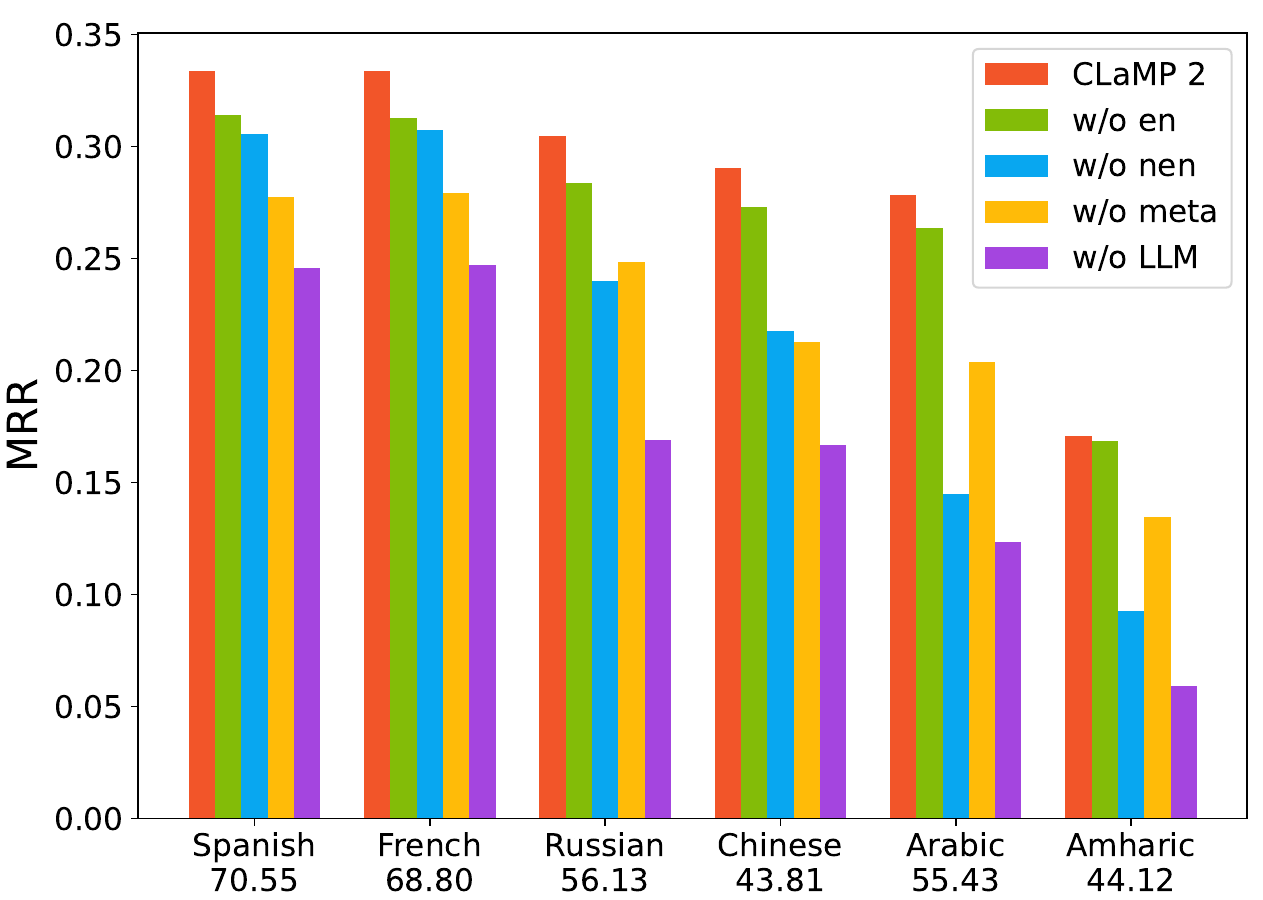}
        \caption{MRR scores on the WikiMT benchmark.}
    \end{subfigure}
    \hfill
    \begin{subfigure}{0.455\textwidth}
        \centering
        \includegraphics[width=\textwidth]{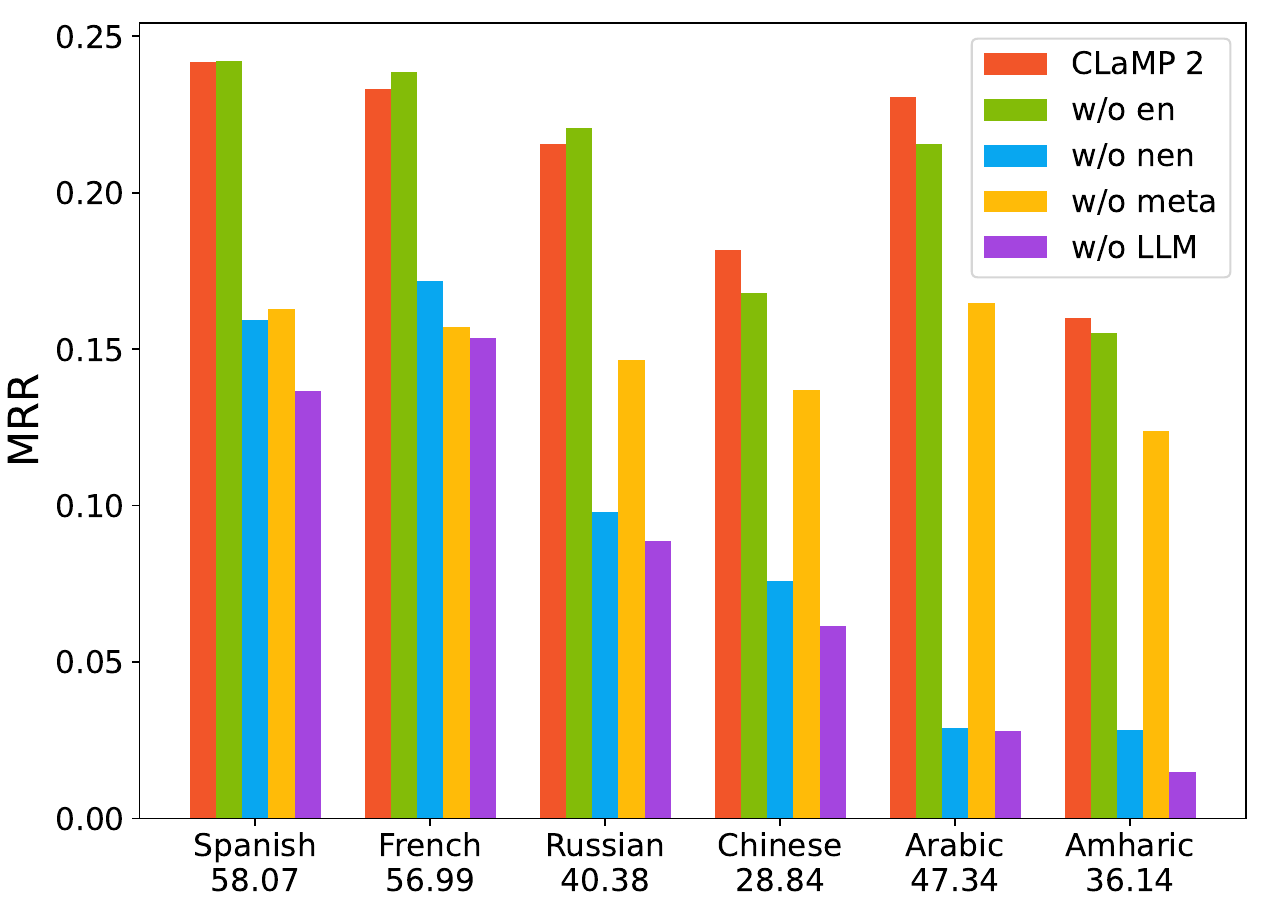}
        \caption{MRR scores on the MidiCaps benchmark.}
    \end{subfigure}
    \caption{MRR scores across six non-English languages for (a) WikiMT and (b) MidiCaps benchmarks. BLEU scores below each language provide additional context on translation quality.}
    \vspace{-2em}
\end{figure}

Table 2 shows semantic search results on the WikiMT and MidiCaps benchmarks, using Mean Reciprocal Rank (MRR) and Hit Rate at Top K (HR@K) to assess model performance in retrieving and ranking relevant music-text pairs.

In the WikiMT benchmark, a clear trend is observed: any CLaMP 2 variant using LLM-generated text, whether in English or non-English, outperforms CLaMP 2 (w/o LLM). For example, CLaMP 2 achieves an MRR of 0.3438. However, when excluding LLM-generated text in CLaMP 2 (w/o LLM), the MRR drops significantly to 0.2797. This indicates that LLM-generated text greatly enhances the CLaMP 2's ability to capture and convey cultural information.

In the MidiCaps benchmark, CLaMP 2 achieves an MRR of 0.2695, demonstrating strong performance. Notably, all CLaMP 2 variants significantly outperform CLaMP. This improvement arises from their native support for MIDI data, enabling a better capture of performance details. In contrast to the WikiMT results, excluding LLM-generated text does not harm performance, as CLaMP 2 (w/o LLM) achieves the highest MRR of 0.2772. This suggests that LLM-generated text may not enhance the understanding of musical features.

\subsection{Semantic Search Across Multilingual Data}
To address the lack of multilingual music-text benchmarks, we translated English texts from WikiMT and MidiCaps into six languages: Spanish, French, Russian, Chinese, Arabic, and Amharic. Among these languages, Amharic is an extremely low-resource language with limited pre-training data—less than 1GB in XLM-R and only 17KB in CLaMP 2. We used SeamlessM4T \cite{DBLP:journals/corr/abs-2308-11596} for its broad translation support, allowing evaluation without native multilingual datasets. Given that translation quality directly affects retrieval effectiveness, we used BLEU scores\footnote{\url{https://github.com/mjpost/sacrebleu}} to assess the similarity between back-translations and original texts, serving as an indicator of translation quality. This evaluation lacks baselines, as no comparable models support multilingual symbolic MIR.

CLaMP 2's multilingual retrieval results are presented in Fig. 5. Generally, removing LLM-generated English texts (w/o en) slightly impacts performance. Although in English, they improve overall text quality by reducing inconsistencies and irrelevancies, thereby enhancing multilingual retrieval performance. In contrast, excluding LLM-generated non-English texts (w/o nen) notably hinders retrieval for all languages in both benchmarks, especially for low-resource languages like Amharic. Comparing CLaMP 2 (w/o en) with CLaMP 2 (w/o LLM) further confirms the important role of LLM-generated non-English texts in enhancing multilingual retrieval performance.

Notably, CLaMP 2 (w/o LLM) records the lowest MRR across all languages in both benchmarks, indicating poor multilingual performance when relying solely on the original text data. However, excluding the original text data (w/o meta) results in a significant drop in performance. This indicates that CLaMP 2 can effectively extract authentic musical concepts from English-centric text data, enabling it to transcend language barriers and improve retrieval across different languages and cultures.

In conclusion, the evaluation of CLaMP 2 on WikiMT and MidiCaps reveals that LLM-generated texts, particularly non-English texts, significantly enhance multilingual semantic search. However, relying solely on them is insufficient, as the original data provides authentic details that LLM-generated data may lack. Together, they enable CLaMP 2 to perform better across languages by learning a more comprehensive representation of music semantics.

\section{Conclusions}
CLaMP 2 makes substantial progress in cross-modal MIR by integrating multilingual text and multimodal music data via contrastive learning. Leveraging GPT-4 to refine the multilingual corpus, it overcomes the limitations of existing models that are exclusively trained on English music-text datasets. This facilitates more precise alignment between music and text across 101 languages.

Experimental results demonstrate that CLaMP 2 achieves state-of-the-art performance across a variety of MIR tasks. In music classification tasks, the M3 model, trained on both ABC and MIDI data, demonstrates improved performance and consistently outperforms counterparts trained on a single modality. Building on M3, CLaMP 2 achieves superior performance across diverse benchmarks and modalities. Notably, the incorporation of LLM-generated text data significantly enhances multilingual semantic search. This enhancement is achieved by reducing textual noise and balancing language distribution, which is particularly beneficial for low-resource languages.

CLaMP 2 establishes a new multilingual MIR standard, enabling users worldwide to access a diverse array of musical content across 101 languages. Future developments may build on CLaMP 2 to connect with audio and visual modalities, facilitating a more comprehensive and culturally rich experience at a global scale.

\section{Excluded Approaches}
In CLaMP 2, several experimental strategies were tested, yet failed to achieve expected improvements and were thus excluded from the final model. It should be noted that these failed attempts are derived from our practice and may not be generalized.

The integration of discretized audio tokens \cite{DBLP:journals/tmlr/DefossezCSA23} failed to match previous audio models' performance and was removed. Inspired by MidiCaps \cite{DBLP:journals/corr/abs-2406-02255} and MuseCoco \cite{DBLP:journals/corr/abs-2306-00110}, we attempted to include musical attributes in the text data. However, this inclusion negatively impacted performance. Additionally, extending the patch masking pre-training strategy to contrastive learning did not enhance the robustness of CLaMP 2. L2 normalization caused convergence problems and was also excluded. Lastly, a learnable logit scale led to over-scaling and degraded representations, so a fixed logit scale of 1 was used for better stability.

\section{Limitations}
Although CLaMP 2 has made progress, it still has certain limitations.

In CLaMP 2, the contrastive learning framework primarily extracts global semantic features, resulting in a loss of fine-grained temporal information. Consequently, tasks that rely on sequential or time-related details cannot be effectively executed.

In addition, the absence of multilingual music-text benchmarks complicates the evaluation of CLaMP 2's performance in non-English languages. To address this, an existing machine translation model \cite{DBLP:journals/corr/abs-2308-11596} was used to translate English benchmarks into other languages. However, machine translation presents its own challenges. For instance, the BLEU score for MidiCaps translations in Chinese is only 28.84, indicating poor translation quality and significantly hindering retrieval performance. Notably, Arabic—despite having far less training data than Chinese in both XLM-R and CLaMP 2—achieves a higher MRR, with a BLEU score of 47.34. This suggests that translation quality has a significant impact on retrieval performance, outweighing the influence of training data size. Without native, high-quality benchmarks for non-English languages, it remains unclear how well CLaMP 2 will perform in real-world multilingual retrieval tasks.

\section*{Core Contributors}
Shangda Wu\textsuperscript{1}, \textit{shangda@mail.ccom.edu.cn}

\noindent
Yashan Wang\textsuperscript{1}, \textit{alexis\_wang@mail.ccom.edu.cn}

\noindent
Ruibin Yuan\textsuperscript{2}, \textit{ryuanab@connect.ust.hk}

\section*{Contributors}
Zhancheng Guo\textsuperscript{1}

\noindent
Xu Tan\textsuperscript{3}

\noindent
Ge Zhang\textsuperscript{2}

\noindent
Monan Zhou\textsuperscript{1}

\noindent
Jing Chen\textsuperscript{4}

\noindent
Xuefeng Mu\textsuperscript{4}

\noindent
Yuejie Gao\textsuperscript{4}

\noindent
Yuanliang Dong\textsuperscript{1}

\noindent
Jiafeng Liu\textsuperscript{1}

\noindent
Xiaobing Li\textsuperscript{1}

\noindent
Feng Yu\textsuperscript{1}

\section*{Correspondence}
Maosong Sun\textsuperscript{1}, \textit{sms@tsinghua.edu.cn}

\section*{Affiliations}
\textsuperscript{1}Central Conservatory of Music, China \\ 
\textsuperscript{2}Multimodal Art Projection Research Community \\ 
\textsuperscript{3}Microsoft Research Asia \\ 
\textsuperscript{4}NetEase Cloud Music, China
    
\section*{Acknowlegdements}
This work was supported by the following funding sources: Special Program of National Natural Science Foundation of China (Grant No. T2341003), Advanced Discipline Construction Project of Beijing Universities, Major Program of National Social Science Fund of China (Grant No. 21ZD19), and the National Culture and Tourism Technological Innovation Engineering Project (Research and Application of 3D Music).

In addition, we would like to express our gratitude for the use of icons from flaticon\footnote{\url{https://www.flaticon.com/}} in Fig. 1 and Fig. 8.

\bibliography{custom}

\clearpage
\newpage
\appendix
\begin{figure}[t]
    \centering
    \includegraphics[width=0.47\textwidth]{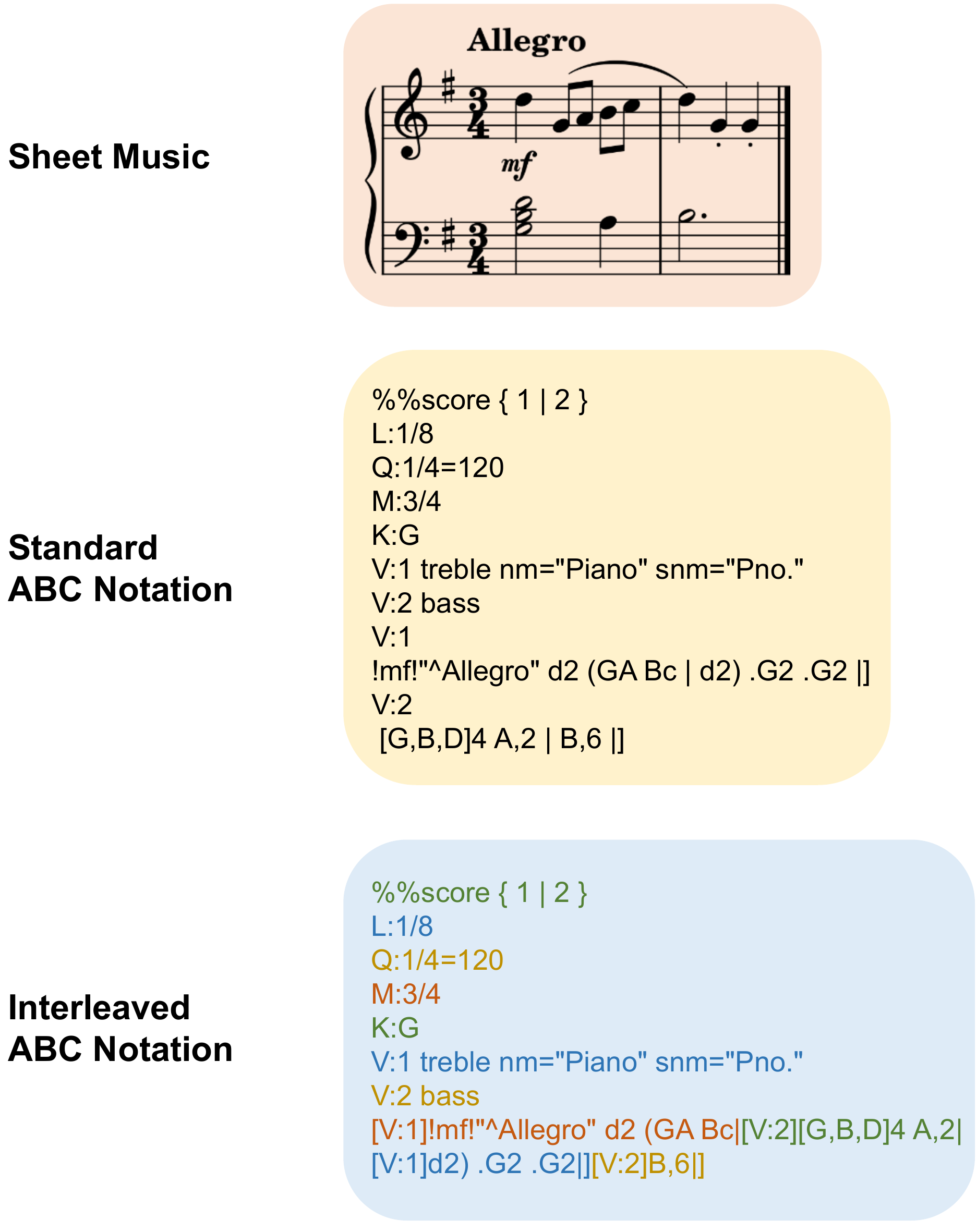}
    \caption{Comparison between standard and interleaved ABC notation in multi-track piano sheet music. Interleaved ABC notation merges voices and tags them in-line for a compact and synchronized representation. Colors mark patch boundaries for M3 model encoding.}
\end{figure}

\section{Interleaved ABC Notation}
Standard ABC notation encodes each voice separately, which often results in corresponding bars being spaced far apart. This separation makes it difficult for models to accurately understand the interactions between voices in sheet music that are meant to align musically.

In contrast, interleaved ABC notation effectively aligns multi-track music by integrating multiple voices of the same bar into a single line, ensuring that all parts remain synchronized. As illustrated in Fig. 6, this format combines voices in-line and tags each bar with its corresponding voice (e.g., \texttt{[V:1]} for treble and \texttt{[V:2]} for bass). By directly aligning related bars, interleaved ABC notation enhances the model’s understanding of how different voices interact within the same bar.

To facilitate this reformatting process, we developed a script for reversible and lossless conversion between standard and interleaved notations, ensuring accuracy without any loss of information. This simplification of multi-track music modeling maintains compatibility with standard ABC syntax, allowing for effective processing in existing tools.

\begin{figure}[t]
    \centering
    \includegraphics[width=0.47\textwidth]{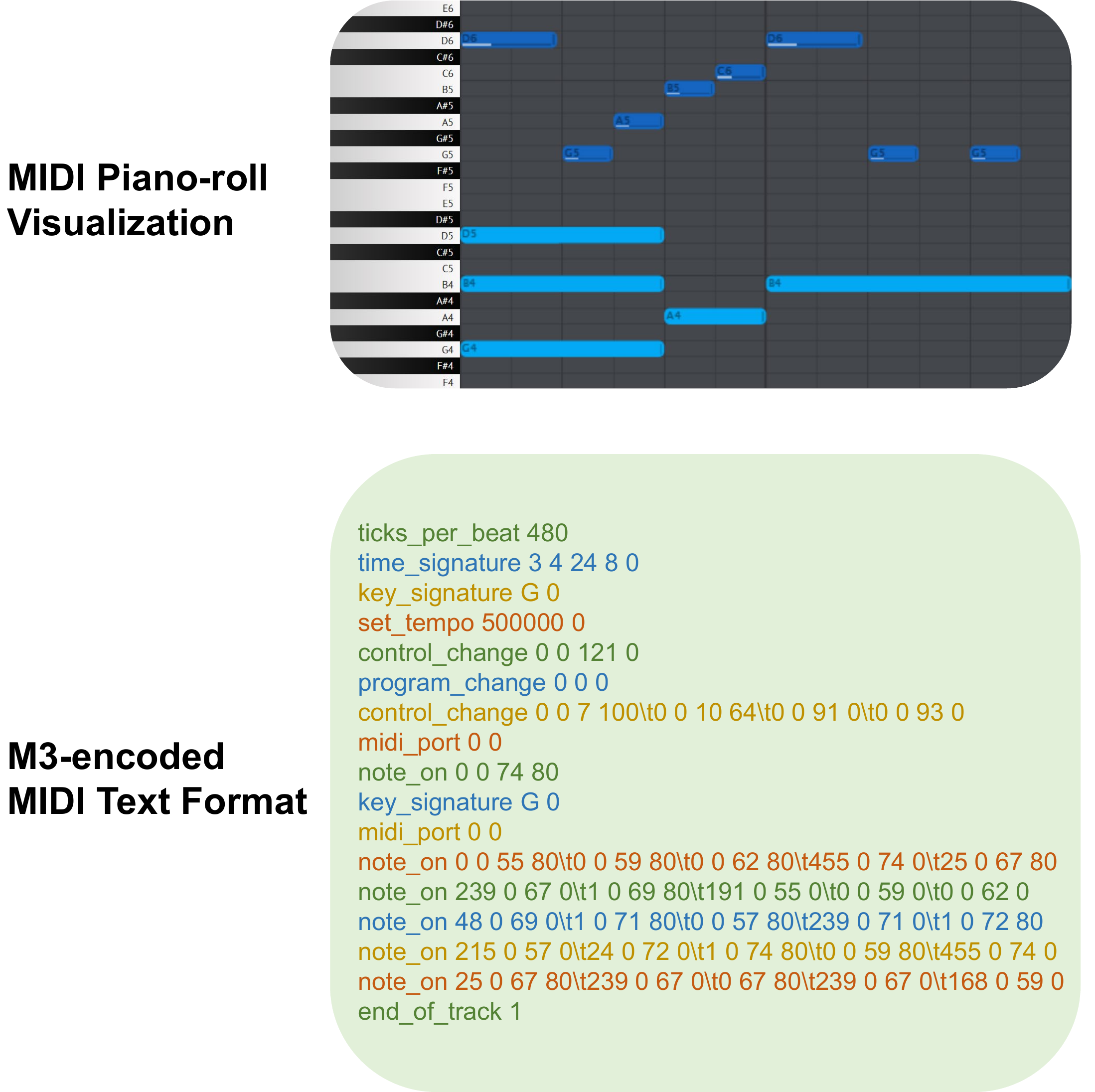}
    \caption{Illustration of MIDI Text Format (MTF) encoded by M3. In this format, MIDI messages are treated as patches for processing. Consecutive messages of the same type are merged within a patch, with colors indicating the boundaries between patches.}
    \vspace{-1em}
\end{figure}

\section{MIDI Text Format}
The MIDI Text Format (MTF) provides a structured, textual representation of MIDI data that preserves all original information without loss. Each MIDI message is accurately represented, allowing full reconstruction from MTF to ensure no musical nuances are overlooked during conversion.

To generate MTF, the mido library reads raw MIDI messages from MIDI files. As shown in Table 3, the output retains all necessary information but can be lengthy and redundant. To simplify this, we streamline the representation by directly reading parameter values in a fixed order and separating them with spaces. For instance, the raw time signature message, which includes multiple parameters—numerator, denominator, clocks per click, notated 32nd notes per beat, and time—is represented in MTF as \texttt{time\_signature 3 4 24 8 0}, as illustrated in Table 4. Other messages, including control changes and note events, are similarly compacted while preserving key musical details.

This approach improves computational performance and maintains precise control of timing and dynamics. Furthermore, when processed by M3, consecutive messages of the same type that fit within a single patch (under 64 characters) are combined into one line, with only the first message containing the type information. This further simplifies representation and improves processing efficiency, as shown in Fig. 7.

\clearpage

\begin{table}[p] 
    \centering
    \caption{Raw MIDI messages extracted from a MIDI file using the mido library.}
    \lstset{basicstyle=\ttfamily\scriptsize,breaklines=true}
    \begin{lstlisting}
MetaMessage('time_signature',
             numerator=3,
             denominator=4,
             clocks_per_click=24,
             notated_32nd_notes_per_beat=8,
             time=0)
MetaMessage('key_signature', key='G', time=0)
MetaMessage('set_tempo', tempo=500000, time=0)
control_change channel=0 control=121 value=0 time=0
program_change channel=0 program=0 time=0
control_change channel=0 control=7 value=100 time=0
control_change channel=0 control=10 value=64 time=0
control_change channel=0 control=91 value=0 time=0
control_change channel=0 control=93 value=0 time=0
MetaMessage('midi_port', port=0, time=0)
note_on channel=0 note=74 velocity=80 time=0
MetaMessage('key_signature', key='G', time=0)
MetaMessage('midi_port', port=0, time=0)
note_on channel=0 note=55 velocity=80 time=0
note_on channel=0 note=59 velocity=80 time=0
note_on channel=0 note=62 velocity=80 time=0
note_on channel=0 note=74 velocity=0 time=455
note_on channel=0 note=67 velocity=80 time=25
note_on channel=0 note=67 velocity=0 time=239
note_on channel=0 note=69 velocity=80 time=1
note_on channel=0 note=55 velocity=0 time=191
note_on channel=0 note=59 velocity=0 time=0
note_on channel=0 note=62 velocity=0 time=0
note_on channel=0 note=69 velocity=0 time=48
note_on channel=0 note=71 velocity=80 time=1
note_on channel=0 note=57 velocity=80 time=0
note_on channel=0 note=71 velocity=0 time=239
note_on channel=0 note=72 velocity=80 time=1
note_on channel=0 note=57 velocity=0 time=215
note_on channel=0 note=72 velocity=0 time=24
note_on channel=0 note=74 velocity=80 time=1
note_on channel=0 note=59 velocity=80 time=0
note_on channel=0 note=74 velocity=0 time=455
note_on channel=0 note=67 velocity=80 time=25
note_on channel=0 note=67 velocity=0 time=239
note_on channel=0 note=67 velocity=80 time=241
note_on channel=0 note=67 velocity=0 time=239
note_on channel=0 note=59 velocity=0 time=168
MetaMessage('end_of_track', time=1)
    \end{lstlisting}
\end{table}

\begin{table}[p] 
    \centering
    \caption{MTF offers a streamlined textual representation of MIDI messages extracted using the mido library. For simplicity, \texttt{ticks\_per\_beat}, though originally an attribute of MIDI objects in mido, is included as the first message at the beginning of the MTF representation.}
    \lstset{basicstyle=\ttfamily\scriptsize,breaklines=true}
    \begin{lstlisting}
ticks_per_beat 480
time_signature 3 4 24 8 0
key_signature G 0
set_tempo 500000 0
control_change 0 0 121 0
program_change 0 0 0
control_change 0 0 7 100
control_change 0 0 10 64
control_change 0 0 91 0
control_change 0 0 93 0
midi_port 0 0
note_on 0 0 74 80
key_signature G 0
midi_port 0 0
note_on 0 0 55 80
note_on 0 0 59 80
note_on 0 0 62 80
note_on 455 0 74 0
note_on 25 0 67 80
note_on 239 0 67 0
note_on 1 0 69 80
note_on 191 0 55 0
note_on 0 0 59 0
note_on 0 0 62 0
note_on 48 0 69 0
note_on 1 0 71 80
note_on 0 0 57 80
note_on 239 0 71 0
note_on 1 0 72 80
note_on 215 0 57 0
note_on 24 0 72 0
note_on 1 0 74 80
note_on 0 0 59 80
note_on 455 0 74 0
note_on 25 0 67 80
note_on 239 0 67 0
note_on 241 0 67 80
note_on 239 0 67 0
note_on 168 0 59 0
end_of_track 1
    \end{lstlisting}
\end{table}

\clearpage
\section{Prompt and Text Examples}
To reduce textual noise and balance language distribution in pre-training data, we carefully designed a structured prompt to leverage the capabilities of GPT-4. As illustrated in Fig. 8, the prompt comprises a system instruction and two conversational examples between a user and the assistant. These examples act as in-context learning references, helping GPT-4 understand the desired output format and the types of information it should extract from the provided metadata.

After formulating the prompt, we organized the metadata entries in our pre-training dataset into a structured JSON format. For each entry, GPT-4 generated corresponding summaries in both English and a randomly selected non-English language from the 99 non-English languages supported by XLM-R \cite{DBLP:conf/acl/ConneauKGCWGGOZ20}, in addition to Cantonese. Including Cantonese, which is well-represented in the dataset and sharing vocabulary with Mandarin, enables CLaMP 2 to support 101 languages without increasing vocabulary size.

To ensure high-quality outputs, both the English and non-English summaries must strictly adhere to the specified JSON format. We implemented filtering criteria to exclude entries that do not meet these requirements, including those returning \texttt{None}, lacking proper JSON structure, or containing non-English summaries in the wrong language. Inconsistencies and structural errors are more prevalent in low-resource languages, as shown in Fig. 4.

To illustrate the effectiveness of this approach, Fig. 9 provides examples that demonstrate GPT-4's ability to generate summaries for different musical compositions. Each example adheres to a structured format, including key metadata—such as the title, composer, genres, description, lyrics, and ensemble information—followed by generated summaries in English and a specified non-English language.

In conclusion, our approach effectively uses GPT-4 to generate structured summaries from noisy, English-centric metadata, reducing textual noise and achieving a more balanced distribution of various languages. By applying filtering criteria, we first remove entries that lack musical information, followed by those that are poorly structured or mismatched with the specified non-English language. This method enhances the quality of our pre-training dataset and promotes a multilingual environment to better serve diverse languages.

\begin{figure}[t]
    \centering
    \includegraphics[width=0.458\textwidth]{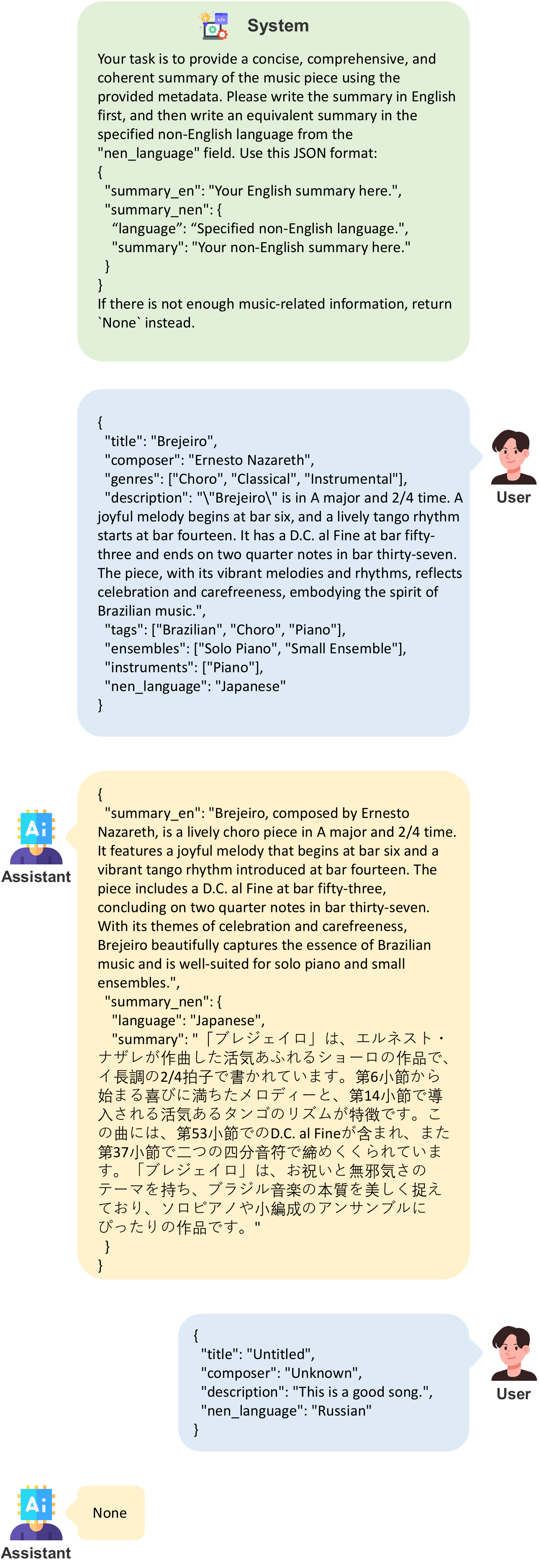}
    \caption{GPT-4 is tasked with generating summaries in English and a selected non-English language. The prompt includes a system instruction and two examples: one shows how to process music metadata—like title, composer, and genre—into clear multilingual summaries, while the other identifies entries lacking sufficient musical information.}
\end{figure}

\begin{figure*}[t]
    \centering
    \includegraphics[width=\textwidth]{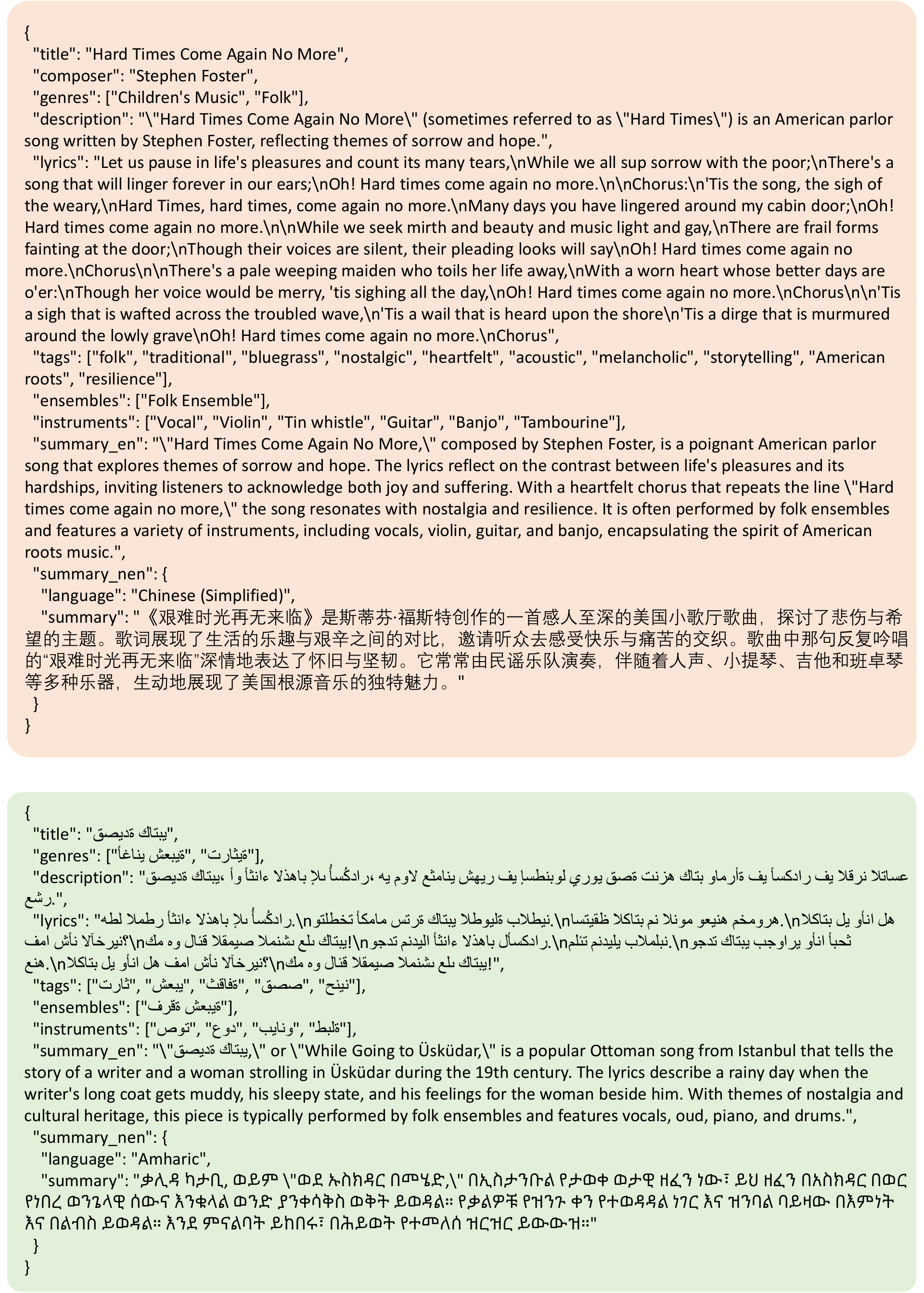}
    \vspace{0.01em}
    \caption{Two examples of LLM-processed text data presented in JSON format, representing the original metadata and LLM-generated summaries in multiple languages for different music pieces.}
\end{figure*}

\clearpage

\section{\textit{t}-SNE Visualizations of CLaMP 2 Representations}
This section presents the \textit{t}-SNE visualizations of feature representations extracted from CLaMP 2 across three benchmarks: WikiMT, VGMIDI, and Pianist8. These visualizations illustrate the clustering patterns of musical representations and reveal an intriguing alignment between the two data modalities—ABC notation and MIDI—without any fine-tuning of the model.

As demonstrated in Fig. 10, the clarity of clustering correlates with the classification performance from Table 1. Pianist8, which achieves the highest accuracy, displays well-defined and tight clusters, signifying that the model adeptly apprehends the minute stylistic subtleties at the composer level.

A particularly notable finding is the mirrored spatial alignment between ABC and MIDI across all datasets. This implies that, despite their dissimilar musical encodings, CLaMP 2 capture comparable latent structures within the feature space. The alignment indicates that CLaMP 2 extracts modality-invariant features—resilient patterns that remain consistent across ABC and MIDI formats. These shared representations are likely to reflect profound musical semantics such as harmonic progressions, rhythmic architectures, or stylistic themes.

This symmetry has practical consequences. It suggests that CLaMP 2 could enable cross-modal tasks, for example, retrieving MIDI files based on ABC queries, without requiring specialized adaptation. It also points to the potential for transfer learning between modalities, where a model trained on one format (e.g., ABC) could operate effectively on another (e.g., MIDI). Future work could explore whether introducing explicit alignment techniques, like contrastive learning, could further enhance cross-modal performance between these two modalities.

These results spotlight both the strengths and limitations of CLaMP 2: the model demonstrates strong generalization across datasets, capturing meaningful musical patterns across diverse domains. However, it struggles with tasks involving overlapping or ambiguous genre boundaries, similar to human perception, such as distinguishing between entities like Bethel and Hillsong, which it finds very similar. This suggests that while CLaMP 2 excels at identifying clear stylistic differences, it may have difficulty differentiating between more closely related or subtle variations.

\begin{figure}[t]
    \centering
    \begin{subfigure}{0.48\textwidth}
        \centering
        \includegraphics[width=\textwidth]{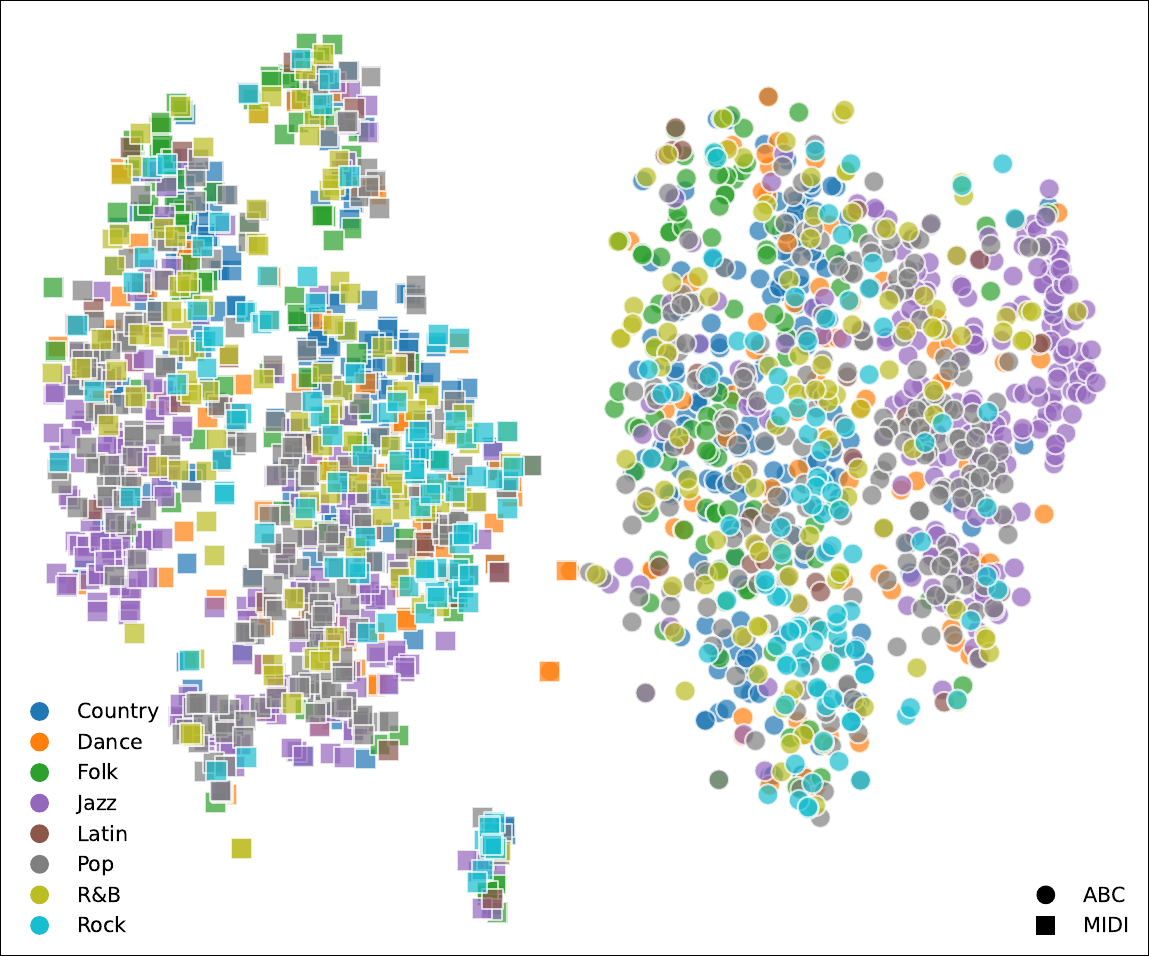}
        \caption{\textit{t}-SNE visualizations on the WikiMT benchmark}
        \vspace{1em}
    \end{subfigure}
    \hfill
    \begin{subfigure}{0.48\textwidth}
        \centering
        \includegraphics[width=\textwidth]{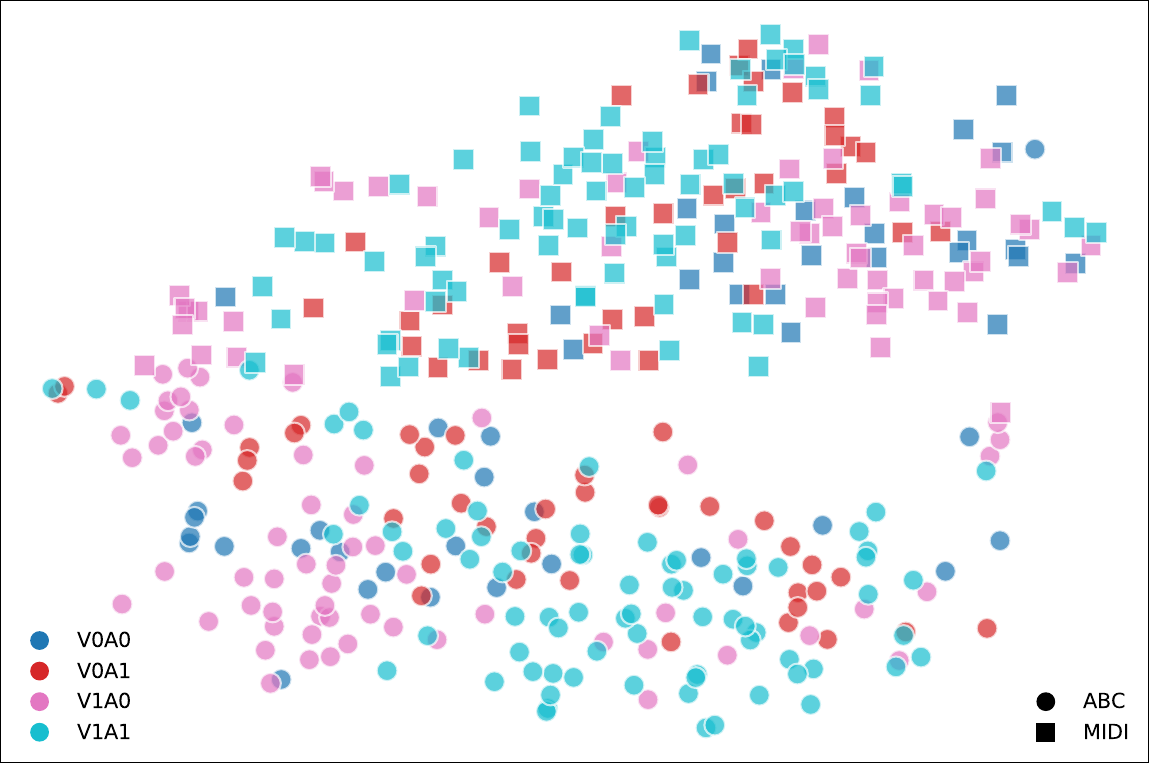}
        \caption{\textit{t}-SNE visualizations on the VGMIDI benchmark}
        \vspace{1em}
    \end{subfigure}
    \vspace{1em}
    \begin{subfigure}{0.48\textwidth}
        \centering
        \includegraphics[width=\textwidth]{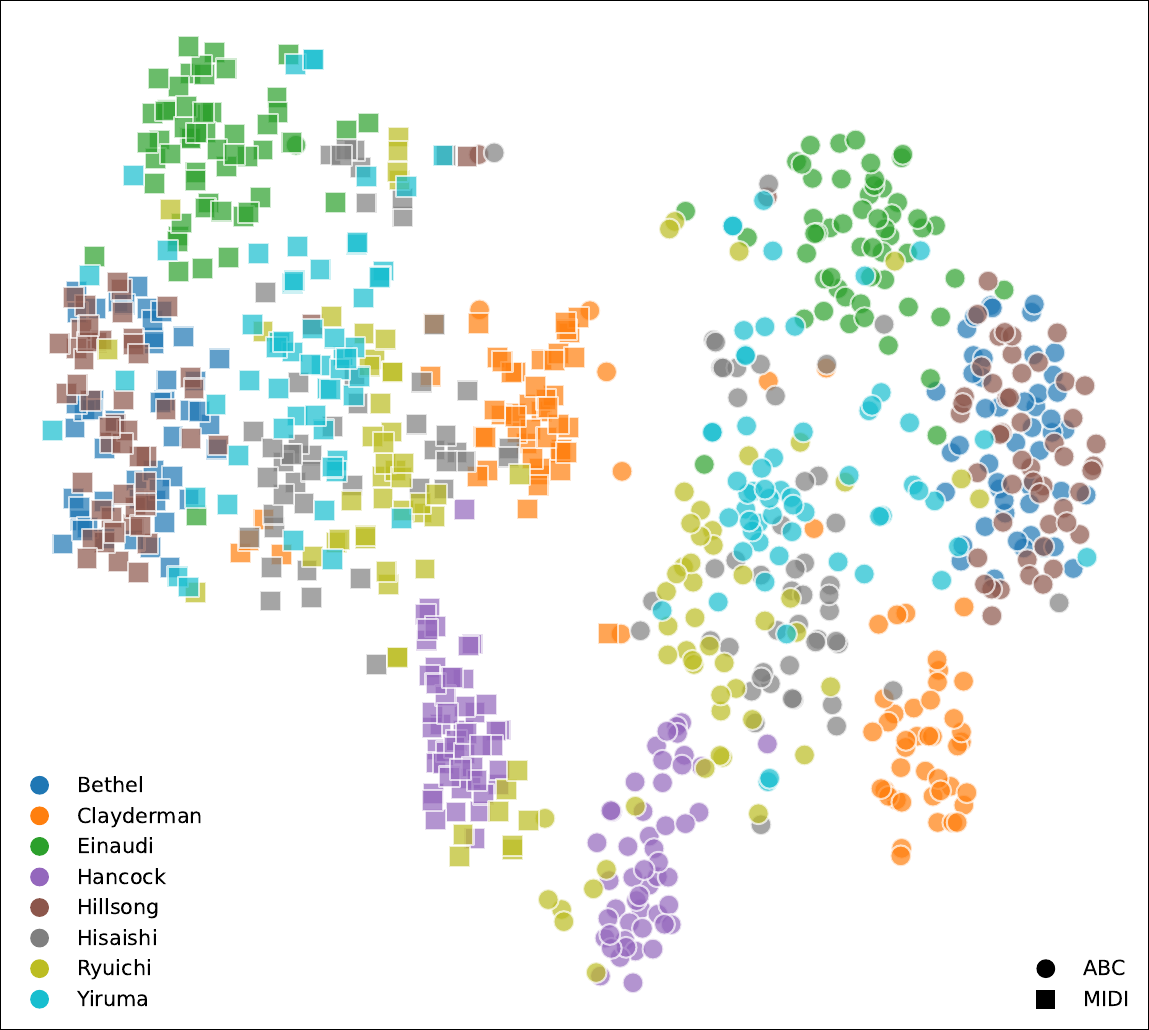}
        \caption{\textit{t}-SNE visualizations on the Pianist8 benchmark}
    \end{subfigure}
    \caption{\textit{t}-SNE visualizations of feature representations from CLaMP 2 (without fine-tuning) for three datasets: (a) WikiMT, (b) VGMIDI, and (c) Pianist8. A noteworthy observation is the mirrored spatial alignment of the ABC and MIDI representations, suggesting that CLaMP 2 effectively extracts modality-invariant musical semantics from both formats.}
    \vspace{-0.2em}
\end{figure}

\end{document}